\documentclass[aps,twocolumn,superscriptaddress,longbibliography,floatfix]{revtex4-1}
\usepackage[colorlinks=true,allcolors=blue, bookmarks=false]{hyperref}
\usepackage{amsfonts,amsmath,amssymb}
\usepackage{graphicx}

\DeclareMathOperator{\Tr}{Tr}

\begin{document}

\title{Gumbel statistics for entanglement spectra of many-body localized eigenstates}

\author{Wouter Buijsman} 

\email{w.buijsman@uva.nl}

\affiliation{Institute for Theoretical Physics Amsterdam and Delta Institute for Theoretical Physics, University of Amsterdam, P.O. Box 94485, 1090 GL Amsterdam, The Netherlands}

\author{Vladimir Gritsev}

\affiliation{Institute for Theoretical Physics Amsterdam and Delta Institute for Theoretical Physics, University of Amsterdam, P.O. Box 94485, 1090 GL Amsterdam, The Netherlands}

\affiliation{Russian Quantum Center, Skolkovo, Moscow 143025, Russia}

\author{Vadim Cheianov}

\affiliation{Instituut-Lorentz and Delta Institute for Theoretical Physics, Universiteit Leiden, P.O. Box 9506, 2300 RA Leiden, The Netherlands}

\date{October 8, 2019}

\begin{abstract}
An entanglement spectrum encodes statistics beyond the entanglement entropy, of which several have been studied in the context of many-body localization. We numerically study the extreme value statistics of entanglement spectra of many-body localized eigenstates. The physical information encoded in these spectra is almost fully carried by the few smallest elements, suggesting the extreme value statistics to have physical significance. We report the surprising observation of Gumbel statistics. Our result provides an analytical, parameter-free characterization of many-body localized eigenstates.
\end{abstract}

\maketitle

\section{Introduction}
Many-body localization (MBL) is understood as a distinct phase of matter that can not be described by conventional statistical physics \cite{Nandkishore15}. Driven by theoretical and experimental progress, there has been a surge of interest in the phenomenon over the last decade \cite{Abanin19}. MBL appears in sufficiently strongly disordered interacting quantum many-body systems, where the appearance of local integrals of motion \cite{Serbyn13-2, Ros15} leads to e.g. emergent integrability \cite{Huse14, Imbrie16}, the absence of thermalization \cite{Serbyn14}, and logarithmic growth of entanglement entropy in time after a quantum quench \cite{Znidaric08, Bardarson12, Serbyn13}.

Thermal and many-body localized phases are separated by an MBL transition \cite{Pal10, Luitz15, Khemani17}. At the localized side of the transition, eigenstates obey area-law scaling of entanglement entropy, while volume-law scaling is observed at the thermal side \cite{Dumitrescu17}. Entanglement entropies can be extracted from entanglement spectra \cite{Li08, Laflorencie16}. An entanglement spectrum encodes statistics beyond the entanglement entropy \cite{Vivo16}, of which several have been studied in the context of MBL \cite{Yang15, Geraedts16, Serbyn16, Yang17, Pietracaprina17, Geraedts17, Gray18}. 

The physical information encoded in the entanglement spectrum of a many-body localized eigenstate is almost fully carried by only a few elements, independent of system size \cite{Serbyn16}. This indicates the potential physical significance of the extreme value statistics \cite{Leadbetter83} of entanglement spectra in the context of MBL. In this work, we study the extreme value statistics of entanglement spectra of many-body localized eigenstates.

Extreme value statistics display universal characteristics over a wide range of physically relevant conditions \cite{Bramwell00, Antal01, Bertin05, Lakshminarayan08}. We report the surprising observation of Gumbel statistics \cite{Gumbel58, Leadbetter83}. These statistics, being observed in studies on various physical phenomena \cite{Bramwell01, Majumdar04, Katzgraber05, Hofferberth08, Lovas17}, apply to the extreme value of $n \to \infty$ independent samples drawn from a distribution with a faster than power-law asymptotic decay. Our result provides an analytical, parameter-free characterization of many-body localized eigenstates.

\section{Gumbel statistics}
Following parts of Ref. \cite{Leadbetter83}, we here briefly discuss Gumbel statistics. Let $X_i$ ($i=1,2,\ldots, n$) be a sequence of independent and identically distributed random variables drawn from a distribution for which the distribution function (the probability that $X_i \le x$) is given by $F(x)$,
\begin{equation}
P \{ X_i \le x \} = F(x).
\end{equation}
Let $M_n$ denote the largest element of the sequence. It follows from the independence of the $X_i$ that the distribution function of $M_n$ is given by
\begin{equation}
P \{ M_n \le x \} = F^n(x).
\end{equation}
Two distribution functions $F_1$ and $F_2$ are said to be of the same type if, up to normalization,
\begin{equation}
F_2(x) = F_1(a x+b)
\label{eq: similar}
\end{equation}
for some $a > 0$ and $b$. From the extremal types theorem, it follows that if
\begin{equation}
\lim_{n \to \infty} F^n(a_n x + b_n) = G(x)
\end{equation}
for some $a_n$ and $b_n$, then $G$ is a distribution function of the same type as one of the three extreme value distributions. The distribution function of one of these extreme value distributions, relevant in the context of this work, is given by
\begin{equation}
G(x) = \exp \left( -e^{-x} \right).
\label{eq: G}
\end{equation}
This type emerges e.g. for a density function $f = dF/dx$ asymptotically decaying faster than a power-law, i.e. as
\begin{equation}
f(x) \sim \exp(-x^\alpha)
\label{eq: tail}
\end{equation}
with $\alpha > 0$ a free parameter. Eq. \eqref{eq: tail} covers e.g. exponential ($\alpha = 1$) and Gaussian ($\alpha = 2$) decays.

The statistics of $G$ given in Eq. \eqref{eq: G} are referred to as Fisher-Tippett-Gumbel \cite{Antal01} or Gumbel \cite{Bertin05} statistics. For these statistics, the rate of convergence depends non-trivially on the structure of $F$ \cite{Gyorgyi10}. When comparing the statistics of a collection of extreme values with Gumbel statistics, it is convenient \cite{Bramwell00, Bramwell01, Antal01, Majumdar04, Katzgraber05, Bertin05, Lakshminarayan08, Hofferberth08, Lovas17} to take $a_n$ and $b_n$ such that the distribution has mean $0$ and standard deviation $1$. The distribution function of the same type as $G$ given in Eq. \eqref{eq: G} with these values for the mean and standard deviation is obtained through Eq. \eqref{eq: similar} with
\begin{equation}
a = \pi / \sqrt{6} , \qquad b = \gamma,
\end{equation}
where $\gamma \approx 0.577$ is Euler's constant. The corresponding density function is given explcitly in Eq. \eqref{eq: dens} below.

\section{Entanglement spectra}
Here, we review the concept of entanglement spectra. In the most general form, the setting is a quantum system divided into subsystems $A$ and $B$ with Hilbert space dimensions $m$ and $n$. A pure state $| \psi \rangle$ of the composite system can be expanded in basis states $| a_i \rangle$ and $| b_i \rangle$ of the respective subsystems as
\begin{equation}
| \psi \rangle = \sum_{i,j} X_{ij} \, | a_i \rangle \otimes | b_j \rangle,
\label{eq: X}
\end{equation}
where $X$ is an $m \times n$ matrix. Labeling the subsystems such that $m \ge n$, the Schmidt decomposition of $X$ uniquely expands $| \psi \rangle$ as a linear combination of product states over the subsystems,
\begin{equation}
| \psi \rangle = \sum_{i=1}^n \sqrt{\lambda_i} \; | \alpha_i \rangle \otimes | \beta_i \rangle,
\end{equation}
where $| \alpha_i \rangle$ and $| \beta_i \rangle$ are basis states for respectively subsystems $A$ and $B$, and the $\lambda_i$ ($\lambda_i \ge 0$) are the Schmidt coefficients. An element $\lambda_i$ can be interpreted as the physical weight of the product state $| \alpha_i \rangle \otimes | \beta_i \rangle$, providing a contribution of $-\lambda_i \ln(\lambda_i)$ to the entanglement entropy. The elements $e_i$ of the entanglement spectrum \cite{Li08, Laflorencie16} are given by
\begin{equation}
e_i = -\ln(\lambda_i).
\end{equation}
We remark that in some literature (e.g. Refs. \cite{Yang15, Serbyn16, Yang17}) the `quantum information definition' $e_i = \lambda_i$ of the entanglement spectrum is used. The smallest of the $e_i$ carry the largest physical weight. In this work, the focus is on the statistics of the smallest of the $e_i$.

The Schmidt coefficients for ergodic (`random') states \cite{Page93} obey the eigenvalue statistics of the fixed-trace Wishart-Laguerre random matrix ensemble \cite{Forrester10}. For this ensemble, the joint density function $ P \{\lambda_{1,2,\ldots, n} = x_{1,2,\ldots,n} \}$ of the eigenvalues is proportional to
\begin{equation}
\prod_{i=1}^n x_i^{\alpha \beta/2} \; \prod_{j<k} |x_j - x_k|^\beta \; \delta \left( \sum_{i=1}^n x_i - 1 \right),
\end{equation}
where $\alpha = (1+m-n)-2 / \beta$, and $\beta$ is the Dyson index given by $1$ or $2$ if the eigenstate is for a system with or without time-reversal symmetry, respectively. The elements are strongly correlated, due to which Gumbel statistics do not apply. For the values of $n$ relevant in the context of this work, the extreme value statistics of the smallest $e_i = -\ln(\lambda_i)$ are close to Gaussian up to $\sim 3$ standard deviations around the mean value (verified numerically).

\section{Physical setting}
We study the eigenstates of a spin chain with random onsite disorder. Let $\sigma^\alpha_i$ denote a Pauli matrix ($\alpha = x,y,z$) acting on site $i$, and let $S^\alpha_i = \sigma^\alpha_i / 2$ denote the corresponding spin-$1/2$ operator. The Hamiltonian of the model under consideration is given by
\begin{equation}
H = \sum_{i=1}^L \left( \vec{S}_i \cdot \vec{S}_{i+1} + h_i S_i^z + \Gamma S_i^x \right).
\label{eq: H}
\end{equation}
We impose periodic boundary conditions $S^\alpha_{L+1} \equiv S^\alpha_{1}$, set $\Gamma = 0.1$, and sample $h_i$ from the uniform distribution ranging over $[-W,W]$. We restrict the focus to eigenstates associated with eigenvalues close to the middle of the spectrum (quantified below) for system sizes $L=10, 12, 14$. The model is studied in Ref. \cite{Yang15}, where indications for an MBL transition at $W \approx 3.5$ are reported. 

The numerical analysis for systems of size $L=10$ or $L=12$ involves the $10$ eigenstates associated with energies closest to the middle $(\max(E_i) + \min(E_i))/2$ of the spectrum $E_i$ ($i = 1,2,\ldots, \dim(H)$), while for systems of size $L=14$ this number is set to $50$. Histograms are drawn from the data of at least $2.5 \times 10^5$ eigenstates, which corresponds to at least $25.000$ disorder realizations for $L=10$ and $L=12$, or at least $5.000$ disorder realizations for $L=14$.

For the calculation of entanglement spectra, we split the chain into subsystems $A$ and $B$ covering respectively the first and last $L/2$ sites, such that $n = 2^{L/2}$. Note that for $\Gamma = 0$ the Hamiltononian reduces to the `standard model of MBL' \cite{Oganesyan07, Pal10, Luitz15}. Then, the total spin projection
\begin{equation}
S^z = \sum_{i=1}^L S^z_i
\end{equation}
is a conserved quantity, due to which the entanglement spectrum is given by the union of independent subspectra labeled by either
\begin{equation}
S^z_A = \sum_{i \in A} S^z_i \qquad \text{or} \qquad S^z_B = \sum_{i \in B} S^z_i.
\end{equation}
This phenomenon is reflected in e.g. a block-diagonal structure of $X$ in Eq. \eqref{eq: X}.

\section{Results}
We here show the main result, namely the observation of Gumbel statistics for the entanglement spectra of many-body localized eigenstates. Let $e_\text{min} = \min_i(e_i)$ denote the smallest element of an entanglement spectrum. Because Gumbel statistics are formulated for the largest element of a sequence, we study the statistics of $-e_\text{min}$.

Let $\langle \cdot \rangle$ denote an expectation value, and let
\begin{equation}
\mu = \langle -e_\text{min} \rangle, \qquad \sigma^2 = \langle e_\text{min}^2 \rangle - \langle e_\text{min} \rangle^2
\end{equation}
denote respectively the mean and variance of the distribution of $-e_\text{min}$. We define $\tilde{e}_\text{min}$ as
\begin{equation}
\tilde{e}_\text{min} = \frac{-e_\text{min} - \mu}{\sqrt{\sigma^2}}.
\end{equation}
By construction, the distribution of $\tilde{e}_\text{min}$ has mean $0$ and standard deviation $1$. We compare the density $P \{ \tilde{e}_\text{min} = x \}$ of $\tilde{e}_\text{min}$ with the density function
\begin{equation}
g(x) = \frac{\pi}{\sqrt{6}} \exp \left[- \left( \frac{\pi}{\sqrt{6}} x + \gamma \right) - e^{-\left( \frac{\pi}{\sqrt{6}} x + \gamma \right)} \right]
\label{eq: dens}
\end{equation}
for Gumbel statistics having the same mean and standard deviation. For reference, we also compare it with the standard Gaussian, approximating the statistics of $\tilde{e}_\text{min}$ for ergodic states, for which the density function is given by
\begin{equation}
h(x) = \frac{1}{\sqrt{2 \pi}} \exp \left( -\frac{1}{2} x^2 \right).
\end{equation}

\begin{figure}
\includegraphics[width= 0.8\columnwidth]{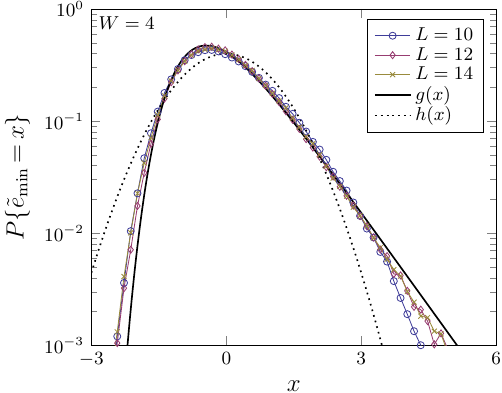}

\bigskip

\includegraphics[width= 0.8\columnwidth]{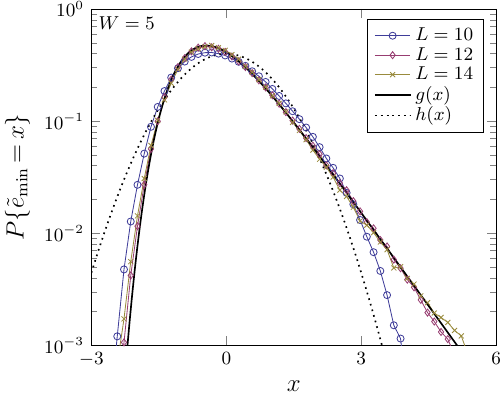}
\caption{Density of $\tilde{e}_\text{min}$ for $W=4$ (top) and $W=5$ (bottom) at $L=10,12,14$, combined with the densities $g(x)$ and $h(x)$. Note the logarithmic scales on the vertical axes.}
\label{fig: Gumbel-W45}
\end{figure}

Fig. \ref{fig: Gumbel-W45} compares the density of $\tilde{e}_\text{min}$ for $W=4$ and $W=5$ at $L=10,12,14$ with $g(x)$ and $h(x)$. The density of $\tilde{e}_\text{min}$ is approximated by a histogram with bins of width $0.05$, which is normalized to unit area to allow for a direct comparison with the (normalized) probability densities. We observe good agreement with $g(x)$ for both disorder strengths at $L=12,14$. Deviations from Gumbel statistics can presumably be attributed to finite-size effects. These effects play a role in both the physics of the eigenstates (becoming stronger localized with increasing system size), as well as in the approach towards the limit $n \to \infty$ (required to observe Gumbel statistics).

\begin{figure}
\includegraphics[width= 0.8\columnwidth]{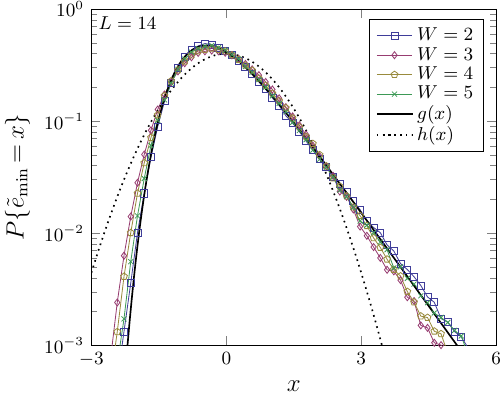} 
\caption{Density of $\tilde{e}_\text{min}$ for $W=2,3,4,5$ at $L=14$, combined with the densities $g(x)$ and $h(x)$. Note the logarithmic scale on the vertical axis.}
\label{fig: Gumbel-L14}
\end{figure}

Fig. \ref{fig: Gumbel-L14} compares the density of $\tilde{e}_\text{min}$ with $g(x)$ and $h(x)$ for $W=2,3,4,5$ at $L=14$. Qualitative similarities between the density of $\tilde{e}_\text{min}$ and $g(x)$ can be observed at all disorder strengths. The eigenstate entanglement spectra of Hamiltonian \eqref{eq: H} are known to show statistics deviating from the expectation for ergodic states already at disorder strengths well below the MBL transition \cite{Yang15}. More generally, the thermal side of the MBL transition is not fully ergodic \cite{Luitz17}. We were not able to draw conclusions on the convergence of the distribution of $\tilde{e}_\text{min}$ towards Gumbel statistics with incrasing system size at the thermal side of the MBL transition, and remark that the statistics of $\tilde{e}_\text{min}$ can not be used as a probe for the MBL transition for the system sizes under consideration. Note that at the thermal side of the MBL transition the physical significance of the statistics of $\tilde{e}_\text{min}$ is presumably limited due to the vanishing physical weight in the thermodynamic limit $L \to \infty$.

\section{Discussion}
\begin{figure}
\includegraphics[width= 0.8\columnwidth]{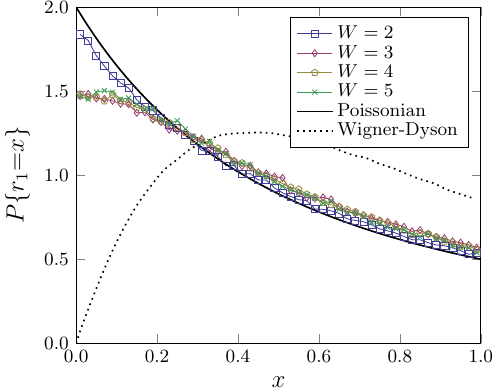} 
\caption{Densities of $r_1$ for $W=2,3,4,5$ at $L=14$, combined with the densities for Poissonian and Wigner-Dyson statistics ($\beta = 1$).}
\label{fig: spacing-L14}
\end{figure}

The observation of Gumbel statistics suggests that the largest elements of an entanglement spectrum $\tilde{e}_i$ are uncorrelated. To verify this, we study the presence of short-range correlations between the largest $\tilde{e}_i$. Short-range correlations can be probed through spacing statistics \cite{Atas13}. We order the $\tilde{e}_i$ in decreasing order (i.e. $\tilde{e}_i \ge \tilde{e}_{i+1}$), and focus on the ratio of consecutive spacings $r_1 \in [0,1]$ given by
\begin{equation}
r _1= \min \left(\frac{\tilde{e}_1 - \tilde{e}_2}{\tilde{e}_2 - \tilde{e}_3}, \frac{\tilde{e}_2 - \tilde{e}_3}{\tilde{e}_1 - \tilde{e}_2} \right).
\end{equation}
In the absence of short-range correlations, the distribution of $r_1$ obeys Poissonian statistics, for which
\begin{equation}
P \{ r_1 = x \} = \frac{2}{(1+x)^2}.
\end{equation}
Ergodic states obey Wigner-Dyson spacing statistics \cite{Mehta04}. For systems with time-reversal symmetry (Dyson index $\beta = 1$), the corresponding density of $r_1$ is well approximated \cite{Atas13} by
\begin{equation}
P \{ r_1 = x \} \approx \frac{27}{8} \frac{x+x^2}{(1+x+x^2)^{5/2}}.
\end{equation}
Fig. \ref{fig: spacing-L14} compares the density of $r_1$ for $W=2,3,4,5$ at $L=14$ with Poissonian and Wigner-Dyson spacing statistics. At all disorder strengths, the spacing statistics are close to Poissonian, indicating the (near) independence of the largest $\tilde{e}_i$.

\section{Conclusion and outlook}
In summary, we have provided numerical evidence that the entanglement spectra of many-body localized eigenstates obey Gumbel statistics. Because the physical weight these spectra is almost fully carried by the few smallest elements, one might expect the extreme value statistics to have physical significance. Our result provides an analytical, parameter-free characterization of many-body localized eigenstates. We stress that no conclusions on the thermal side of the MBL transition can be drawn.

The main open question remaining is the physical mechanism responsible for the occurence of Gumbel statistics, and the way it can be explained in terms of phenomenological models \cite{Huse14}. A possible starting point for further investigations might be provided by the notion that an entanglement spectrum can be interpreted as the eigenvalue spectrum of the entanglement Hamiltonian
\begin{equation}
H_\text{ent} = -\ln(\rho_B),
\end{equation}
where $\rho_B = \Tr_A(| \psi \rangle \langle \psi |)$ is the partial trace of the density matrix $| \Psi \rangle \langle \Psi |$ over basis states of subsystem $A$ \cite{Li08}. One might hypothesize that the statistics of the eigenvector associated with $e_\text{min}$ carry relevant information.

\begin{acknowledgments}
We thank Maksym Serbyn for useful discussions. This work is part of the Delta-ITP consortium, a program of the Netherlands Organization for Scientific Research (NWO) that is funded by the Dutch Ministry of Education, Culture and Science (OCW).
\end{acknowledgments}

\bibliography{Gumbel}

\begin{thebibliography}{42}%
\makeatletter
\providecommand \@ifxundefined [1]{%
 \@ifx{#1\undefined}
}%
\providecommand \@ifnum [1]{%
 \ifnum #1\expandafter \@firstoftwo
 \else \expandafter \@secondoftwo
 \fi
}%
\providecommand \@ifx [1]{%
 \ifx #1\expandafter \@firstoftwo
 \else \expandafter \@secondoftwo
 \fi
}%
\providecommand \natexlab [1]{#1}%
\providecommand \enquote  [1]{``#1''}%
\providecommand \bibnamefont  [1]{#1}%
\providecommand \bibfnamefont [1]{#1}%
\providecommand \citenamefont [1]{#1}%
\providecommand \href@noop [0]{\@secondoftwo}%
\providecommand \href [0]{\begingroup \@sanitize@url \@href}%
\providecommand \@href[1]{\@@startlink{#1}\@@href}%
\providecommand \@@href[1]{\endgroup#1\@@endlink}%
\providecommand \@sanitize@url [0]{\catcode `\\12\catcode `\$12\catcode
  `\&12\catcode `\#12\catcode `\^12\catcode `\_12\catcode `\%12\relax}%
\providecommand \@@startlink[1]{}%
\providecommand \@@endlink[0]{}%
\providecommand \url  [0]{\begingroup\@sanitize@url \@url }%
\providecommand \@url [1]{\endgroup\@href {#1}{\urlprefix }}%
\providecommand \urlprefix  [0]{URL }%
\providecommand \Eprint [0]{\href }%
\providecommand \doibase [0]{http://dx.doi.org/}%
\providecommand \selectlanguage [0]{\@gobble}%
\providecommand \bibinfo  [0]{\@secondoftwo}%
\providecommand \bibfield  [0]{\@secondoftwo}%
\providecommand \translation [1]{[#1]}%
\providecommand \BibitemOpen [0]{}%
\providecommand \bibitemStop [0]{}%
\providecommand \bibitemNoStop [0]{.\EOS\space}%
\providecommand \EOS [0]{\spacefactor3000\relax}%
\providecommand \BibitemShut  [1]{\csname bibitem#1\endcsname}%
\let\auto@bib@innerbib\@empty
\bibitem [{\citenamefont {Nandkishore}\ and\ \citenamefont
  {Huse}(2015)}]{Nandkishore15}%
  \BibitemOpen
  \bibfield  {author} {\bibinfo {author} {\bibfnamefont {R.}~\bibnamefont
  {Nandkishore}}\ and\ \bibinfo {author} {\bibfnamefont {D.~A.}\ \bibnamefont
  {Huse}},\ }\bibfield  {title} {\enquote {\bibinfo {title} {Many-body
  localization and thermalization in quantum statistical mechanics},}\ }\href
  {\doibase 10.1146/annurev-conmatphys-031214-014726} {\bibfield  {journal}
  {\bibinfo  {journal} {Annu. Rev. Condens. Matter Phys}\ }\textbf {\bibinfo
  {volume} {6}},\ \bibinfo {pages} {15} (\bibinfo {year} {2015})}\BibitemShut
  {NoStop}%
\bibitem [{\citenamefont {Abanin}\ \emph {et~al.}(2019)\citenamefont {Abanin},
  \citenamefont {Altman}, \citenamefont {Bloch},\ and\ \citenamefont
  {Serbyn}}]{Abanin19}%
  \BibitemOpen
  \bibfield  {author} {\bibinfo {author} {\bibfnamefont {D.~A.}\ \bibnamefont
  {Abanin}}, \bibinfo {author} {\bibfnamefont {E.}~\bibnamefont {Altman}},
  \bibinfo {author} {\bibfnamefont {I.}~\bibnamefont {Bloch}}, \ and\ \bibinfo
  {author} {\bibfnamefont {M.}~\bibnamefont {Serbyn}},\ }\bibfield  {title}
  {\enquote {\bibinfo {title} {Colloquium: {Many-body} localization,
  thermalization, and entanglement},}\ }\href {\doibase
  10.1103/RevModPhys.91.021001} {\bibfield  {journal} {\bibinfo  {journal}
  {Rev. Mod. Phys.}\ }\textbf {\bibinfo {volume} {91}},\ \bibinfo {pages}
  {021001} (\bibinfo {year} {2019})}\BibitemShut {NoStop}%
\bibitem [{\citenamefont {Serbyn}\ \emph
  {et~al.}(2013{\natexlab{a}})\citenamefont {Serbyn}, \citenamefont
  {Papi\ifmmode~\acute{c}\else \'{c}\fi{}},\ and\ \citenamefont
  {Abanin}}]{Serbyn13-2}%
  \BibitemOpen
  \bibfield  {author} {\bibinfo {author} {\bibfnamefont {M.}~\bibnamefont
  {Serbyn}}, \bibinfo {author} {\bibfnamefont {Z.}~\bibnamefont
  {Papi\ifmmode~\acute{c}\else \'{c}\fi{}}}, \ and\ \bibinfo {author}
  {\bibfnamefont {D.~A.}\ \bibnamefont {Abanin}},\ }\bibfield  {title}
  {\enquote {\bibinfo {title} {Local {Conservation} {Laws} and the {Structure}
  of the {Many-Body} {Localized} {States}},}\ }\href {\doibase
  10.1103/PhysRevLett.111.127201} {\bibfield  {journal} {\bibinfo  {journal}
  {Phys. Rev. Lett.}\ }\textbf {\bibinfo {volume} {111}},\ \bibinfo {pages}
  {127201} (\bibinfo {year} {2013}{\natexlab{a}})}\BibitemShut {NoStop}%
\bibitem [{\citenamefont {Ros}\ \emph {et~al.}(2015)\citenamefont {Ros},
  \citenamefont {M\"uller},\ and\ \citenamefont {Scardicchio}}]{Ros15}%
  \BibitemOpen
  \bibfield  {author} {\bibinfo {author} {\bibfnamefont {V.}~\bibnamefont
  {Ros}}, \bibinfo {author} {\bibfnamefont {M.}~\bibnamefont {M\"uller}}, \
  and\ \bibinfo {author} {\bibfnamefont {A.}~\bibnamefont {Scardicchio}},\
  }\bibfield  {title} {\enquote {\bibinfo {title} {Integrals of motion in the
  many-body localized phase},}\ }\href {\doibase
  10.1016/j.nuclphysb.2014.12.014} {\bibfield  {journal} {\bibinfo  {journal}
  {Nucl. Phys. B}\ }\textbf {\bibinfo {volume} {891}},\ \bibinfo {pages} {420}
  (\bibinfo {year} {2015})}\BibitemShut {NoStop}%
\bibitem [{\citenamefont {Huse}\ \emph {et~al.}(2014)\citenamefont {Huse},
  \citenamefont {Nandkishore},\ and\ \citenamefont {Oganesyan}}]{Huse14}%
  \BibitemOpen
  \bibfield  {author} {\bibinfo {author} {\bibfnamefont {D.~A.}\ \bibnamefont
  {Huse}}, \bibinfo {author} {\bibfnamefont {R.}~\bibnamefont {Nandkishore}}, \
  and\ \bibinfo {author} {\bibfnamefont {V.}~\bibnamefont {Oganesyan}},\
  }\bibfield  {title} {\enquote {\bibinfo {title} {Phenomenology of fully
  many-body-localized systems},}\ }\href {\doibase 10.1103/PhysRevB.90.174202}
  {\bibfield  {journal} {\bibinfo  {journal} {Phys. Rev. B}\ }\textbf {\bibinfo
  {volume} {90}},\ \bibinfo {pages} {174202} (\bibinfo {year}
  {2014})}\BibitemShut {NoStop}%
\bibitem [{\citenamefont {Imbrie}(2016)}]{Imbrie16}%
  \BibitemOpen
  \bibfield  {author} {\bibinfo {author} {\bibfnamefont {J.~Z.}\ \bibnamefont
  {Imbrie}},\ }\bibfield  {title} {\enquote {\bibinfo {title} {Diagonalization
  and {Many-Body} {Localization} for a {Disordered} {Quantum} {Spin}
  {Chain}},}\ }\href {\doibase 10.1103/PhysRevLett.117.027201} {\bibfield
  {journal} {\bibinfo  {journal} {Phys. Rev. Lett.}\ }\textbf {\bibinfo
  {volume} {117}},\ \bibinfo {pages} {027201} (\bibinfo {year}
  {2016})}\BibitemShut {NoStop}%
\bibitem [{\citenamefont {Serbyn}\ \emph {et~al.}(2014)\citenamefont {Serbyn},
  \citenamefont {Papi\ifmmode~\acute{c}\else \'{c}\fi{}},\ and\ \citenamefont
  {Abanin}}]{Serbyn14}%
  \BibitemOpen
  \bibfield  {author} {\bibinfo {author} {\bibfnamefont {M.}~\bibnamefont
  {Serbyn}}, \bibinfo {author} {\bibfnamefont {Z.}~\bibnamefont
  {Papi\ifmmode~\acute{c}\else \'{c}\fi{}}}, \ and\ \bibinfo {author}
  {\bibfnamefont {D.~A.}\ \bibnamefont {Abanin}},\ }\bibfield  {title}
  {\enquote {\bibinfo {title} {Quantum quenches in the many-body localized
  phase},}\ }\href {\doibase 10.1103/PhysRevB.90.174302} {\bibfield  {journal}
  {\bibinfo  {journal} {Phys. Rev. B}\ }\textbf {\bibinfo {volume} {90}},\
  \bibinfo {pages} {174302} (\bibinfo {year} {2014})}\BibitemShut {NoStop}%
\bibitem [{\citenamefont {\ifmmode \check{Z}\else
  \v{Z}\fi{}nidari\ifmmode~\check{c}\else \v{c}\fi{}}\ \emph
  {et~al.}(2008)\citenamefont {\ifmmode \check{Z}\else
  \v{Z}\fi{}nidari\ifmmode~\check{c}\else \v{c}\fi{}}, \citenamefont {Prosen},\
  and\ \citenamefont {Prelov\ifmmode~\check{s}\else
  \v{s}\fi{}ek}}]{Znidaric08}%
  \BibitemOpen
  \bibfield  {author} {\bibinfo {author} {\bibfnamefont {M.}~\bibnamefont
  {\ifmmode \check{Z}\else \v{Z}\fi{}nidari\ifmmode~\check{c}\else
  \v{c}\fi{}}}, \bibinfo {author} {\bibfnamefont {T.}~\bibnamefont {Prosen}}, \
  and\ \bibinfo {author} {\bibfnamefont {P.}~\bibnamefont
  {Prelov\ifmmode~\check{s}\else \v{s}\fi{}ek}},\ }\bibfield  {title} {\enquote
  {\bibinfo {title} {Many-body localization in the {Heisenberg} $xxz$ magnet in
  a random field},}\ }\href {\doibase 10.1103/PhysRevB.77.064426} {\bibfield
  {journal} {\bibinfo  {journal} {Phys. Rev. B}\ }\textbf {\bibinfo {volume}
  {77}},\ \bibinfo {pages} {064426} (\bibinfo {year} {2008})}\BibitemShut
  {NoStop}%
\bibitem [{\citenamefont {Bardarson}\ \emph {et~al.}(2012)\citenamefont
  {Bardarson}, \citenamefont {Pollmann},\ and\ \citenamefont
  {Moore}}]{Bardarson12}%
  \BibitemOpen
  \bibfield  {author} {\bibinfo {author} {\bibfnamefont {J.~H.}\ \bibnamefont
  {Bardarson}}, \bibinfo {author} {\bibfnamefont {F.}~\bibnamefont {Pollmann}},
  \ and\ \bibinfo {author} {\bibfnamefont {J.~E.}\ \bibnamefont {Moore}},\
  }\bibfield  {title} {\enquote {\bibinfo {title} {Unbounded {Growth} of
  {Entanglement} in {Models} of {Many-Body} {Localization}},}\ }\href {\doibase
  10.1103/PhysRevLett.109.017202} {\bibfield  {journal} {\bibinfo  {journal}
  {Phys. Rev. Lett.}\ }\textbf {\bibinfo {volume} {109}},\ \bibinfo {pages}
  {017202} (\bibinfo {year} {2012})}\BibitemShut {NoStop}%
\bibitem [{\citenamefont {Serbyn}\ \emph
  {et~al.}(2013{\natexlab{b}})\citenamefont {Serbyn}, \citenamefont
  {Papi\ifmmode~\acute{c}\else \'{c}\fi{}},\ and\ \citenamefont
  {Abanin}}]{Serbyn13}%
  \BibitemOpen
  \bibfield  {author} {\bibinfo {author} {\bibfnamefont {M.}~\bibnamefont
  {Serbyn}}, \bibinfo {author} {\bibfnamefont {Z.}~\bibnamefont
  {Papi\ifmmode~\acute{c}\else \'{c}\fi{}}}, \ and\ \bibinfo {author}
  {\bibfnamefont {D.~A.}\ \bibnamefont {Abanin}},\ }\bibfield  {title}
  {\enquote {\bibinfo {title} {Universal {Slow} {Growth} of {Entanglement} in
  {Interacting} {Strongly} {Disordered} {Systems}},}\ }\href {\doibase
  10.1103/PhysRevLett.110.260601} {\bibfield  {journal} {\bibinfo  {journal}
  {Phys. Rev. Lett.}\ }\textbf {\bibinfo {volume} {110}},\ \bibinfo {pages}
  {260601} (\bibinfo {year} {2013}{\natexlab{b}})}\BibitemShut {NoStop}%
\bibitem [{\citenamefont {Pal}\ and\ \citenamefont {Huse}(2010)}]{Pal10}%
  \BibitemOpen
  \bibfield  {author} {\bibinfo {author} {\bibfnamefont {A.}~\bibnamefont
  {Pal}}\ and\ \bibinfo {author} {\bibfnamefont {D.~A.}\ \bibnamefont {Huse}},\
  }\bibfield  {title} {\enquote {\bibinfo {title} {Many-body localization phase
  transition},}\ }\href {\doibase 10.1103/PhysRevB.82.174411} {\bibfield
  {journal} {\bibinfo  {journal} {Phys. Rev. B}\ }\textbf {\bibinfo {volume}
  {82}},\ \bibinfo {pages} {174411} (\bibinfo {year} {2010})}\BibitemShut
  {NoStop}%
\bibitem [{\citenamefont {Luitz}\ \emph {et~al.}(2015)\citenamefont {Luitz},
  \citenamefont {Laflorencie},\ and\ \citenamefont {Alet}}]{Luitz15}%
  \BibitemOpen
  \bibfield  {author} {\bibinfo {author} {\bibfnamefont {D.~J.}\ \bibnamefont
  {Luitz}}, \bibinfo {author} {\bibfnamefont {N.}~\bibnamefont {Laflorencie}},
  \ and\ \bibinfo {author} {\bibfnamefont {F.}~\bibnamefont {Alet}},\
  }\bibfield  {title} {\enquote {\bibinfo {title} {Many-body localization edge
  in the random-field {Heisenberg} chain},}\ }\href {\doibase
  10.1103/PhysRevB.91.081103} {\bibfield  {journal} {\bibinfo  {journal} {Phys.
  Rev. B}\ }\textbf {\bibinfo {volume} {91}},\ \bibinfo {pages} {081103(R)}
  (\bibinfo {year} {2015})}\BibitemShut {NoStop}%
\bibitem [{\citenamefont {Khemani}\ \emph {et~al.}(2017)\citenamefont
  {Khemani}, \citenamefont {Lim}, \citenamefont {Sheng},\ and\ \citenamefont
  {Huse}}]{Khemani17}%
  \BibitemOpen
  \bibfield  {author} {\bibinfo {author} {\bibfnamefont {V.}~\bibnamefont
  {Khemani}}, \bibinfo {author} {\bibfnamefont {S.~P.}\ \bibnamefont {Lim}},
  \bibinfo {author} {\bibfnamefont {D.~N.}\ \bibnamefont {Sheng}}, \ and\
  \bibinfo {author} {\bibfnamefont {D.~A.}\ \bibnamefont {Huse}},\ }\bibfield
  {title} {\enquote {\bibinfo {title} {Critical {Properties} of the {Many-Body}
  {Localization} {Transition}},}\ }\href {\doibase 10.1103/PhysRevX.7.021013}
  {\bibfield  {journal} {\bibinfo  {journal} {Phys. Rev. X}\ }\textbf {\bibinfo
  {volume} {7}},\ \bibinfo {pages} {021013} (\bibinfo {year}
  {2017})}\BibitemShut {NoStop}%
\bibitem [{\citenamefont {Dumitrescu}\ \emph {et~al.}(2017)\citenamefont
  {Dumitrescu}, \citenamefont {Vasseur},\ and\ \citenamefont
  {Potter}}]{Dumitrescu17}%
  \BibitemOpen
  \bibfield  {author} {\bibinfo {author} {\bibfnamefont {P.~T.}\ \bibnamefont
  {Dumitrescu}}, \bibinfo {author} {\bibfnamefont {R.}~\bibnamefont {Vasseur}},
  \ and\ \bibinfo {author} {\bibfnamefont {A.~C.}\ \bibnamefont {Potter}},\
  }\bibfield  {title} {\enquote {\bibinfo {title} {Scaling {Theory} of
  {Entanglement} at the {Many-Body} {Localization} {Transition}},}\ }\href
  {\doibase 10.1103/PhysRevLett.119.110604} {\bibfield  {journal} {\bibinfo
  {journal} {Phys. Rev. Lett.}\ }\textbf {\bibinfo {volume} {119}},\ \bibinfo
  {pages} {110604} (\bibinfo {year} {2017})}\BibitemShut {NoStop}%
\bibitem [{\citenamefont {Li}\ and\ \citenamefont {Haldane}(2008)}]{Li08}%
  \BibitemOpen
  \bibfield  {author} {\bibinfo {author} {\bibfnamefont {H.}~\bibnamefont
  {Li}}\ and\ \bibinfo {author} {\bibfnamefont {F.~D.~M.}\ \bibnamefont
  {Haldane}},\ }\bibfield  {title} {\enquote {\bibinfo {title} {Entanglement
  {Spectrum} as a {Generalization} of {Entanglement} {Entropy}:
  {Identification} of {Topological} {Order} in {Non-Abelian} {Fractional}
  {Quantum} {Hall} {Effect} {States}},}\ }\href {\doibase
  10.1103/PhysRevLett.101.010504} {\bibfield  {journal} {\bibinfo  {journal}
  {Phys. Rev. Lett.}\ }\textbf {\bibinfo {volume} {101}},\ \bibinfo {pages}
  {010504} (\bibinfo {year} {2008})}\BibitemShut {NoStop}%
\bibitem [{\citenamefont {Laflorencie}(2016)}]{Laflorencie16}%
  \BibitemOpen
  \bibfield  {author} {\bibinfo {author} {\bibfnamefont {N.}~\bibnamefont
  {Laflorencie}},\ }\bibfield  {title} {\enquote {\bibinfo {title} {Quantum
  entanglement in condensed matter systems},}\ }\href {\doibase
  10.1016/j.physrep.2016.06.008} {\bibfield  {journal} {\bibinfo  {journal}
  {Phys. Rep.}\ }\textbf {\bibinfo {volume} {646}},\ \bibinfo {pages} {1}
  (\bibinfo {year} {2016})}\BibitemShut {NoStop}%
\bibitem [{\citenamefont {Vivo}\ \emph {et~al.}(2016)\citenamefont {Vivo},
  \citenamefont {Pato},\ and\ \citenamefont {Oshanin}}]{Vivo16}%
  \BibitemOpen
  \bibfield  {author} {\bibinfo {author} {\bibfnamefont {P.}~\bibnamefont
  {Vivo}}, \bibinfo {author} {\bibfnamefont {M.~P.}\ \bibnamefont {Pato}}, \
  and\ \bibinfo {author} {\bibfnamefont {G.}~\bibnamefont {Oshanin}},\
  }\bibfield  {title} {\enquote {\bibinfo {title} {Random pure states:
  {Quantifying} bipartite entanglement beyond the linear statistics},}\ }\href
  {\doibase 10.1103/PhysRevE.93.052106} {\bibfield  {journal} {\bibinfo
  {journal} {Phys. Rev. E}\ }\textbf {\bibinfo {volume} {93}},\ \bibinfo
  {pages} {052106} (\bibinfo {year} {2016})}\BibitemShut {NoStop}%
\bibitem [{\citenamefont {Yang}\ \emph {et~al.}(2015)\citenamefont {Yang},
  \citenamefont {Chamon}, \citenamefont {Hamma},\ and\ \citenamefont
  {Mucciolo}}]{Yang15}%
  \BibitemOpen
  \bibfield  {author} {\bibinfo {author} {\bibfnamefont {Z.-C.}\ \bibnamefont
  {Yang}}, \bibinfo {author} {\bibfnamefont {C.}~\bibnamefont {Chamon}},
  \bibinfo {author} {\bibfnamefont {A.}~\bibnamefont {Hamma}}, \ and\ \bibinfo
  {author} {\bibfnamefont {E.~R.}\ \bibnamefont {Mucciolo}},\ }\bibfield
  {title} {\enquote {\bibinfo {title} {{Two-Component} {Structure} in the
  {Entanglement} {Spectrum} of {Highly} {Excited} {States}},}\ }\href {\doibase
  10.1103/PhysRevLett.115.267206} {\bibfield  {journal} {\bibinfo  {journal}
  {Phys. Rev. Lett.}\ }\textbf {\bibinfo {volume} {115}},\ \bibinfo {pages}
  {267206} (\bibinfo {year} {2015})}\BibitemShut {NoStop}%
\bibitem [{\citenamefont {Geraedts}\ \emph {et~al.}(2016)\citenamefont
  {Geraedts}, \citenamefont {Nandkishore},\ and\ \citenamefont
  {Regnault}}]{Geraedts16}%
  \BibitemOpen
  \bibfield  {author} {\bibinfo {author} {\bibfnamefont {S.~D.}\ \bibnamefont
  {Geraedts}}, \bibinfo {author} {\bibfnamefont {R.}~\bibnamefont
  {Nandkishore}}, \ and\ \bibinfo {author} {\bibfnamefont {N.}~\bibnamefont
  {Regnault}},\ }\bibfield  {title} {\enquote {\bibinfo {title} {Many-body
  localization and thermalization: {Insights} from the entanglement
  spectrum},}\ }\href {\doibase 10.1103/PhysRevB.93.174202} {\bibfield
  {journal} {\bibinfo  {journal} {Phys. Rev. B}\ }\textbf {\bibinfo {volume}
  {93}},\ \bibinfo {pages} {174202} (\bibinfo {year} {2016})}\BibitemShut
  {NoStop}%
\bibitem [{\citenamefont {Serbyn}\ \emph {et~al.}(2016)\citenamefont {Serbyn},
  \citenamefont {Michailidis}, \citenamefont {Abanin},\ and\ \citenamefont
  {Papi\ifmmode~\acute{c}\else \'{c}\fi{}}}]{Serbyn16}%
  \BibitemOpen
  \bibfield  {author} {\bibinfo {author} {\bibfnamefont {M.}~\bibnamefont
  {Serbyn}}, \bibinfo {author} {\bibfnamefont {A.~A.}\ \bibnamefont
  {Michailidis}}, \bibinfo {author} {\bibfnamefont {D.~A.}\ \bibnamefont
  {Abanin}}, \ and\ \bibinfo {author} {\bibfnamefont {Z.}~\bibnamefont
  {Papi\ifmmode~\acute{c}\else \'{c}\fi{}}},\ }\bibfield  {title} {\enquote
  {\bibinfo {title} {{Power-Law} {Entanglement} {Spectrum} in {Many-Body}
  {Localized} {Phases}},}\ }\href {\doibase 10.1103/PhysRevLett.117.160601}
  {\bibfield  {journal} {\bibinfo  {journal} {Phys. Rev. Lett.}\ }\textbf
  {\bibinfo {volume} {117}},\ \bibinfo {pages} {160601} (\bibinfo {year}
  {2016})}\BibitemShut {NoStop}%
\bibitem [{\citenamefont {Yang}\ \emph {et~al.}(2017)\citenamefont {Yang},
  \citenamefont {Hamma}, \citenamefont {Giampaolo}, \citenamefont {Mucciolo},\
  and\ \citenamefont {Chamon}}]{Yang17}%
  \BibitemOpen
  \bibfield  {author} {\bibinfo {author} {\bibfnamefont {Z.-C.}\ \bibnamefont
  {Yang}}, \bibinfo {author} {\bibfnamefont {A.}~\bibnamefont {Hamma}},
  \bibinfo {author} {\bibfnamefont {S.~M.}\ \bibnamefont {Giampaolo}}, \bibinfo
  {author} {\bibfnamefont {E.~R.}\ \bibnamefont {Mucciolo}}, \ and\ \bibinfo
  {author} {\bibfnamefont {C.}~\bibnamefont {Chamon}},\ }\bibfield  {title}
  {\enquote {\bibinfo {title} {Entanglement complexity in quantum many-body
  dynamics, thermalization, and localization},}\ }\href {\doibase
  10.1103/PhysRevB.96.020408} {\bibfield  {journal} {\bibinfo  {journal} {Phys.
  Rev. B}\ }\textbf {\bibinfo {volume} {96}},\ \bibinfo {pages} {020408(R)}
  (\bibinfo {year} {2017})}\BibitemShut {NoStop}%
\bibitem [{\citenamefont {Pietracaprina}\ \emph {et~al.}(2017)\citenamefont
  {Pietracaprina}, \citenamefont {Parisi}, \citenamefont {Mariano},
  \citenamefont {Pascazio},\ and\ \citenamefont
  {Scardicchio}}]{Pietracaprina17}%
  \BibitemOpen
  \bibfield  {author} {\bibinfo {author} {\bibfnamefont {F.}~\bibnamefont
  {Pietracaprina}}, \bibinfo {author} {\bibfnamefont {G.}~\bibnamefont
  {Parisi}}, \bibinfo {author} {\bibfnamefont {A.}~\bibnamefont {Mariano}},
  \bibinfo {author} {\bibfnamefont {S.}~\bibnamefont {Pascazio}}, \ and\
  \bibinfo {author} {\bibfnamefont {A.}~\bibnamefont {Scardicchio}},\
  }\bibfield  {title} {\enquote {\bibinfo {title} {Entanglement critical length
  at the many-body localization transition},}\ }\href {\doibase
  10.1088/1742-5468/aa9338} {\bibfield  {journal} {\bibinfo  {journal} {J.
  Stat. Mech.}\ }\textbf {\bibinfo {volume} {2017}},\ \bibinfo {pages} {113102}
  (\bibinfo {year} {2017})}\BibitemShut {NoStop}%
\bibitem [{\citenamefont {Geraedts}\ \emph {et~al.}(2017)\citenamefont
  {Geraedts}, \citenamefont {Regnault},\ and\ \citenamefont
  {Nandkishore}}]{Geraedts17}%
  \BibitemOpen
  \bibfield  {author} {\bibinfo {author} {\bibfnamefont {S.~D.}\ \bibnamefont
  {Geraedts}}, \bibinfo {author} {\bibfnamefont {N.}~\bibnamefont {Regnault}},
  \ and\ \bibinfo {author} {\bibfnamefont {R.~M.}\ \bibnamefont
  {Nandkishore}},\ }\bibfield  {title} {\enquote {\bibinfo {title}
  {Characterizing the many-body localization transition using the entanglement
  spectrum},}\ }\href {\doibase 10.1088/1367-2630/aa93a5} {\bibfield  {journal}
  {\bibinfo  {journal} {New J. Phys.}\ }\textbf {\bibinfo {volume} {19}},\
  \bibinfo {pages} {113021} (\bibinfo {year} {2017})}\BibitemShut {NoStop}%
\bibitem [{\citenamefont {Gray}\ \emph {et~al.}(2018)\citenamefont {Gray},
  \citenamefont {Bose},\ and\ \citenamefont {Bayat}}]{Gray18}%
  \BibitemOpen
  \bibfield  {author} {\bibinfo {author} {\bibfnamefont {J.}~\bibnamefont
  {Gray}}, \bibinfo {author} {\bibfnamefont {S.}~\bibnamefont {Bose}}, \ and\
  \bibinfo {author} {\bibfnamefont {A.}~\bibnamefont {Bayat}},\ }\bibfield
  {title} {\enquote {\bibinfo {title} {Many-body localization transition:
  {Schmidt gap}, entanglement length, and scaling},}\ }\href {\doibase
  10.1103/PhysRevB.97.201105} {\bibfield  {journal} {\bibinfo  {journal} {Phys.
  Rev. B}\ }\textbf {\bibinfo {volume} {97}},\ \bibinfo {pages} {201105(R)}
  (\bibinfo {year} {2018})}\BibitemShut {NoStop}%
\bibitem [{\citenamefont {Leadbetter}\ \emph {et~al.}(1983)\citenamefont
  {Leadbetter}, \citenamefont {Lindgren},\ and\ \citenamefont
  {Rootz\'en}}]{Leadbetter83}%
  \BibitemOpen
  \bibfield  {author} {\bibinfo {author} {\bibfnamefont {M.~R.}\ \bibnamefont
  {Leadbetter}}, \bibinfo {author} {\bibfnamefont {G.}~\bibnamefont
  {Lindgren}}, \ and\ \bibinfo {author} {\bibfnamefont {H.}~\bibnamefont
  {Rootz\'en}},\ }\href@noop {} {\emph {\bibinfo {title} {Extremes and
  {Related} {Properties} of {Random} {Sequences} and {Processes}}}},\ \bibinfo
  {series} {Springer {Series} in {Statistics}}, Vol.~\bibinfo {volume} {11}\
  (\bibinfo  {publisher} {Springer-Verlag},\ \bibinfo {address} {New York,
  Heidelberg, Berlin},\ \bibinfo {year} {1983})\BibitemShut {NoStop}%
\bibitem [{\citenamefont {Bramwell}\ \emph {et~al.}(2000)\citenamefont
  {Bramwell}, \citenamefont {Christensen}, \citenamefont {Fortin},
  \citenamefont {Holdsworth}, \citenamefont {Jensen}, \citenamefont {Lise},
  \citenamefont {L\'opez}, \citenamefont {Nicodemi}, \citenamefont {Pinton},\
  and\ \citenamefont {Sellitto}}]{Bramwell00}%
  \BibitemOpen
  \bibfield  {author} {\bibinfo {author} {\bibfnamefont {S.~T.}\ \bibnamefont
  {Bramwell}}, \bibinfo {author} {\bibfnamefont {K.}~\bibnamefont
  {Christensen}}, \bibinfo {author} {\bibfnamefont {J.-Y.}\ \bibnamefont
  {Fortin}}, \bibinfo {author} {\bibfnamefont {P.~C.~W.}\ \bibnamefont
  {Holdsworth}}, \bibinfo {author} {\bibfnamefont {H.~J.}\ \bibnamefont
  {Jensen}}, \bibinfo {author} {\bibfnamefont {S.}~\bibnamefont {Lise}},
  \bibinfo {author} {\bibfnamefont {J.~M.}\ \bibnamefont {L\'opez}}, \bibinfo
  {author} {\bibfnamefont {M.}~\bibnamefont {Nicodemi}}, \bibinfo {author}
  {\bibfnamefont {J.-F.}\ \bibnamefont {Pinton}}, \ and\ \bibinfo {author}
  {\bibfnamefont {M.}~\bibnamefont {Sellitto}},\ }\bibfield  {title} {\enquote
  {\bibinfo {title} {Universal {Fluctuations} in {Correlated} {Systems}},}\
  }\href {\doibase 10.1103/PhysRevLett.84.3744} {\bibfield  {journal} {\bibinfo
   {journal} {Phys. Rev. Lett.}\ }\textbf {\bibinfo {volume} {84}},\ \bibinfo
  {pages} {3744} (\bibinfo {year} {2000})}\BibitemShut {NoStop}%
\bibitem [{\citenamefont {Antal}\ \emph {et~al.}(2001)\citenamefont {Antal},
  \citenamefont {Droz}, \citenamefont {Gy\"orgyi},\ and\ \citenamefont
  {R\'acz}}]{Antal01}%
  \BibitemOpen
  \bibfield  {author} {\bibinfo {author} {\bibfnamefont {T.}~\bibnamefont
  {Antal}}, \bibinfo {author} {\bibfnamefont {M.}~\bibnamefont {Droz}},
  \bibinfo {author} {\bibfnamefont {G.}~\bibnamefont {Gy\"orgyi}}, \ and\
  \bibinfo {author} {\bibfnamefont {Z.}~\bibnamefont {R\'acz}},\ }\bibfield
  {title} {\enquote {\bibinfo {title} {$1/\mathit{f}$ {Noise} and {Extreme}
  {Value} {Statistics}},}\ }\href {\doibase 10.1103/PhysRevLett.87.240601}
  {\bibfield  {journal} {\bibinfo  {journal} {Phys. Rev. Lett.}\ }\textbf
  {\bibinfo {volume} {87}},\ \bibinfo {pages} {240601} (\bibinfo {year}
  {2001})}\BibitemShut {NoStop}%
\bibitem [{\citenamefont {Bertin}(2005)}]{Bertin05}%
  \BibitemOpen
  \bibfield  {author} {\bibinfo {author} {\bibfnamefont {E.}~\bibnamefont
  {Bertin}},\ }\bibfield  {title} {\enquote {\bibinfo {title} {Global
  {Fluctuations} and {Gumbel} {Statistics}},}\ }\href {\doibase
  10.1103/PhysRevLett.95.170601} {\bibfield  {journal} {\bibinfo  {journal}
  {Phys. Rev. Lett.}\ }\textbf {\bibinfo {volume} {95}},\ \bibinfo {pages}
  {170601} (\bibinfo {year} {2005})}\BibitemShut {NoStop}%
\bibitem [{\citenamefont {Lakshminarayan}\ \emph {et~al.}(2008)\citenamefont
  {Lakshminarayan}, \citenamefont {Tomsovic}, \citenamefont {Bohigas},\ and\
  \citenamefont {Majumdar}}]{Lakshminarayan08}%
  \BibitemOpen
  \bibfield  {author} {\bibinfo {author} {\bibfnamefont {A.}~\bibnamefont
  {Lakshminarayan}}, \bibinfo {author} {\bibfnamefont {S.}~\bibnamefont
  {Tomsovic}}, \bibinfo {author} {\bibfnamefont {O.}~\bibnamefont {Bohigas}}, \
  and\ \bibinfo {author} {\bibfnamefont {S.~N.}\ \bibnamefont {Majumdar}},\
  }\bibfield  {title} {\enquote {\bibinfo {title} {Extreme {Statistics} of
  {Complex} {Random} and {Quantum} {Chaotic} {States}},}\ }\href {\doibase
  10.1103/PhysRevLett.100.044103} {\bibfield  {journal} {\bibinfo  {journal}
  {Phys. Rev. Lett.}\ }\textbf {\bibinfo {volume} {100}},\ \bibinfo {pages}
  {044103} (\bibinfo {year} {2008})}\BibitemShut {NoStop}%
\bibitem [{\citenamefont {Gumbel}(1958)}]{Gumbel58}%
  \BibitemOpen
  \bibfield  {author} {\bibinfo {author} {\bibfnamefont {E.~J.}\ \bibnamefont
  {Gumbel}},\ }\href@noop {} {\emph {\bibinfo {title} {Statistics of
  {Extremes}}}}\ (\bibinfo  {publisher} {Colubia University Press},\ \bibinfo
  {address} {New York},\ \bibinfo {year} {1958})\BibitemShut {NoStop}%
\bibitem [{\citenamefont {Bramwell}\ \emph {et~al.}(2001)\citenamefont
  {Bramwell}, \citenamefont {Fortin}, \citenamefont {Holdsworth}, \citenamefont
  {Peysson}, \citenamefont {Pinton}, \citenamefont {Portelli},\ and\
  \citenamefont {Sellitto}}]{Bramwell01}%
  \BibitemOpen
  \bibfield  {author} {\bibinfo {author} {\bibfnamefont {S.~T.}\ \bibnamefont
  {Bramwell}}, \bibinfo {author} {\bibfnamefont {J.-Y.}\ \bibnamefont
  {Fortin}}, \bibinfo {author} {\bibfnamefont {P.~C.~W.}\ \bibnamefont
  {Holdsworth}}, \bibinfo {author} {\bibfnamefont {S.}~\bibnamefont {Peysson}},
  \bibinfo {author} {\bibfnamefont {J.-F.}\ \bibnamefont {Pinton}}, \bibinfo
  {author} {\bibfnamefont {B.}~\bibnamefont {Portelli}}, \ and\ \bibinfo
  {author} {\bibfnamefont {M.}~\bibnamefont {Sellitto}},\ }\bibfield  {title}
  {\enquote {\bibinfo {title} {Magnetic fluctuations in the classical
  $\mathrm{XY}$ model: The origin of an exponential tail in a complex
  system},}\ }\href {\doibase 10.1103/PhysRevE.63.041106} {\bibfield  {journal}
  {\bibinfo  {journal} {Phys. Rev. E}\ }\textbf {\bibinfo {volume} {63}},\
  \bibinfo {pages} {041106} (\bibinfo {year} {2001})}\BibitemShut {NoStop}%
\bibitem [{\citenamefont {Majumdar}\ and\ \citenamefont
  {Comtet}(2004)}]{Majumdar04}%
  \BibitemOpen
  \bibfield  {author} {\bibinfo {author} {\bibfnamefont {S.~N.}\ \bibnamefont
  {Majumdar}}\ and\ \bibinfo {author} {\bibfnamefont {A.}~\bibnamefont
  {Comtet}},\ }\bibfield  {title} {\enquote {\bibinfo {title} {Exact {Maximal}
  {Height} {Distribution} of {Fluctuating} {Interfaces}},}\ }\href {\doibase
  10.1103/PhysRevLett.92.225501} {\bibfield  {journal} {\bibinfo  {journal}
  {Phys. Rev. Lett.}\ }\textbf {\bibinfo {volume} {92}},\ \bibinfo {pages}
  {225501} (\bibinfo {year} {2004})}\BibitemShut {NoStop}%
\bibitem [{\citenamefont {Katzgraber}\ \emph {et~al.}(2005)\citenamefont
  {Katzgraber}, \citenamefont {K\"orner}, \citenamefont {Liers}, \citenamefont
  {J\"unger},\ and\ \citenamefont {Hartmann}}]{Katzgraber05}%
  \BibitemOpen
  \bibfield  {author} {\bibinfo {author} {\bibfnamefont {H.~G.}\ \bibnamefont
  {Katzgraber}}, \bibinfo {author} {\bibfnamefont {M.}~\bibnamefont
  {K\"orner}}, \bibinfo {author} {\bibfnamefont {F.}~\bibnamefont {Liers}},
  \bibinfo {author} {\bibfnamefont {M.}~\bibnamefont {J\"unger}}, \ and\
  \bibinfo {author} {\bibfnamefont {A.~K.}\ \bibnamefont {Hartmann}},\
  }\bibfield  {title} {\enquote {\bibinfo {title} {Universality-class
  dependence of energy distributions in spin glasses},}\ }\href {\doibase
  10.1103/PhysRevB.72.094421} {\bibfield  {journal} {\bibinfo  {journal} {Phys.
  Rev. B}\ }\textbf {\bibinfo {volume} {72}},\ \bibinfo {pages} {094421}
  (\bibinfo {year} {2005})}\BibitemShut {NoStop}%
\bibitem [{\citenamefont {Hofferberth}\ \emph {et~al.}(2008)\citenamefont
  {Hofferberth}, \citenamefont {Lesanovsky}, \citenamefont {Schumm},
  \citenamefont {Imambekov}, \citenamefont {Gritsev}, \citenamefont {Demler},\
  and\ \citenamefont {Schmiedmayer}}]{Hofferberth08}%
  \BibitemOpen
  \bibfield  {author} {\bibinfo {author} {\bibfnamefont {S.}~\bibnamefont
  {Hofferberth}}, \bibinfo {author} {\bibfnamefont {I.}~\bibnamefont
  {Lesanovsky}}, \bibinfo {author} {\bibfnamefont {T.}~\bibnamefont {Schumm}},
  \bibinfo {author} {\bibfnamefont {A.~I.}\ \bibnamefont {Imambekov}}, \bibinfo
  {author} {\bibfnamefont {V.}~\bibnamefont {Gritsev}}, \bibinfo {author}
  {\bibfnamefont {E.}~\bibnamefont {Demler}}, \ and\ \bibinfo {author}
  {\bibfnamefont {J.}~\bibnamefont {Schmiedmayer}},\ }\bibfield  {title}
  {\enquote {\bibinfo {title} {Probing quantum and thermal noise in an
  interacting many-body system},}\ }\href {\doibase 10.1038/nphys941}
  {\bibfield  {journal} {\bibinfo  {journal} {Nat. Phys.}\ }\textbf {\bibinfo
  {volume} {4}},\ \bibinfo {pages} {489} (\bibinfo {year} {2008})}\BibitemShut
  {NoStop}%
\bibitem [{\citenamefont {Lovas}\ \emph {et~al.}(2017)\citenamefont {Lovas},
  \citenamefont {D\'ora}, \citenamefont {Demler},\ and\ \citenamefont
  {Zar\'and}}]{Lovas17}%
  \BibitemOpen
  \bibfield  {author} {\bibinfo {author} {\bibfnamefont {I.}~\bibnamefont
  {Lovas}}, \bibinfo {author} {\bibfnamefont {B.}~\bibnamefont {D\'ora}},
  \bibinfo {author} {\bibfnamefont {E.}~\bibnamefont {Demler}}, \ and\ \bibinfo
  {author} {\bibfnamefont {G.}~\bibnamefont {Zar\'and}},\ }\bibfield  {title}
  {\enquote {\bibinfo {title} {Full counting statistics of time-of-flight
  images},}\ }\href {\doibase 10.1103/PhysRevA.95.053621} {\bibfield  {journal}
  {\bibinfo  {journal} {Phys. Rev. A}\ }\textbf {\bibinfo {volume} {95}},\
  \bibinfo {pages} {053621} (\bibinfo {year} {2017})}\BibitemShut {NoStop}%
\bibitem [{\citenamefont {Gy\"orgyi}\ \emph {et~al.}(2010)\citenamefont
  {Gy\"orgyi}, \citenamefont {Moloney}, \citenamefont {Ozog\'any},
  \citenamefont {R\'acz},\ and\ \citenamefont {Droz}}]{Gyorgyi10}%
  \BibitemOpen
  \bibfield  {author} {\bibinfo {author} {\bibfnamefont {G.}~\bibnamefont
  {Gy\"orgyi}}, \bibinfo {author} {\bibfnamefont {N.~R.}\ \bibnamefont
  {Moloney}}, \bibinfo {author} {\bibfnamefont {K.}~\bibnamefont {Ozog\'any}},
  \bibinfo {author} {\bibfnamefont {Z.}~\bibnamefont {R\'acz}}, \ and\ \bibinfo
  {author} {\bibfnamefont {M.}~\bibnamefont {Droz}},\ }\bibfield  {title}
  {\enquote {\bibinfo {title} {Renormalization-group theory for finite-size
  scaling in extreme statistics},}\ }\href {\doibase
  10.1103/PhysRevE.81.041135} {\bibfield  {journal} {\bibinfo  {journal} {Phys.
  Rev. E}\ }\textbf {\bibinfo {volume} {81}},\ \bibinfo {pages} {041135}
  (\bibinfo {year} {2010})}\BibitemShut {NoStop}%
\bibitem [{\citenamefont {Page}(1993)}]{Page93}%
  \BibitemOpen
  \bibfield  {author} {\bibinfo {author} {\bibfnamefont {D.~N.}\ \bibnamefont
  {Page}},\ }\bibfield  {title} {\enquote {\bibinfo {title} {Average entropy of
  a subsystem},}\ }\href {\doibase 10.1103/PhysRevLett.71.1291} {\bibfield
  {journal} {\bibinfo  {journal} {Phys. Rev. Lett.}\ }\textbf {\bibinfo
  {volume} {71}},\ \bibinfo {pages} {1291} (\bibinfo {year}
  {1993})}\BibitemShut {NoStop}%
\bibitem [{\citenamefont {Forrester}(2010)}]{Forrester10}%
  \BibitemOpen
  \bibfield  {author} {\bibinfo {author} {\bibfnamefont {P.~J.}\ \bibnamefont
  {Forrester}},\ }\href@noop {} {\emph {\bibinfo {title} {Log-Gases and
  {Random} {Matrices}}}},\ \bibinfo {series} {London {Mathematical} {Society}
  {Monographs}}, Vol.~\bibinfo {volume} {34}\ (\bibinfo  {publisher} {Princeton
  University Press},\ \bibinfo {address} {Princeton and Oxford},\ \bibinfo
  {year} {2010})\BibitemShut {NoStop}%
\bibitem [{\citenamefont {Oganesyan}\ and\ \citenamefont
  {Huse}(2007)}]{Oganesyan07}%
  \BibitemOpen
  \bibfield  {author} {\bibinfo {author} {\bibfnamefont {V.}~\bibnamefont
  {Oganesyan}}\ and\ \bibinfo {author} {\bibfnamefont {D.~A.}\ \bibnamefont
  {Huse}},\ }\bibfield  {title} {\enquote {\bibinfo {title} {Localization of
  interacting fermions at high temperature},}\ }\href {\doibase
  10.1103/PhysRevB.75.155111} {\bibfield  {journal} {\bibinfo  {journal} {Phys.
  Rev. B}\ }\textbf {\bibinfo {volume} {75}},\ \bibinfo {pages} {155111}
  (\bibinfo {year} {2007})}\BibitemShut {NoStop}%
\bibitem [{\citenamefont {Luitz}\ and\ \citenamefont
  {Bar~Lev}(2017)}]{Luitz17}%
  \BibitemOpen
  \bibfield  {author} {\bibinfo {author} {\bibfnamefont {D.~J.}\ \bibnamefont
  {Luitz}}\ and\ \bibinfo {author} {\bibfnamefont {Y.}~\bibnamefont
  {Bar~Lev}},\ }\bibfield  {title} {\enquote {\bibinfo {title} {The ergodic
  side of the many-body localization transition},}\ }\href {\doibase
  10.1002/andp.201600350} {\bibfield  {journal} {\bibinfo  {journal} {Ann.
  Phys.}\ }\textbf {\bibinfo {volume} {529}},\ \bibinfo {pages} {1600350}
  (\bibinfo {year} {2017})}\BibitemShut {NoStop}%
\bibitem [{\citenamefont {Atas}\ \emph {et~al.}(2013)\citenamefont {Atas},
  \citenamefont {Bogomolny}, \citenamefont {Giraud},\ and\ \citenamefont
  {Roux}}]{Atas13}%
  \BibitemOpen
  \bibfield  {author} {\bibinfo {author} {\bibfnamefont {Y.~Y.}\ \bibnamefont
  {Atas}}, \bibinfo {author} {\bibfnamefont {E.}~\bibnamefont {Bogomolny}},
  \bibinfo {author} {\bibfnamefont {O.}~\bibnamefont {Giraud}}, \ and\ \bibinfo
  {author} {\bibfnamefont {G.}~\bibnamefont {Roux}},\ }\bibfield  {title}
  {\enquote {\bibinfo {title} {Distribution of the {Ratio} of {Consecutive}
  {Level} {Spacings} in {Random} {Matrix} {Ensembles}},}\ }\href {\doibase
  10.1103/PhysRevLett.110.084101} {\bibfield  {journal} {\bibinfo  {journal}
  {Phys. Rev. Lett.}\ }\textbf {\bibinfo {volume} {110}},\ \bibinfo {pages}
  {084101} (\bibinfo {year} {2013})}\BibitemShut {NoStop}%
\bibitem [{\citenamefont {Mehta}(2004)}]{Mehta04}%
  \BibitemOpen
  \bibfield  {author} {\bibinfo {author} {\bibfnamefont {M.~L.}\ \bibnamefont
  {Mehta}},\ }\href
  {https://www.sciencedirect.com/bookseries/pure-and-applied-mathematics/vol/142}
  {\emph {\bibinfo {title} {Random {Matrices}}}},\ \bibinfo {edition} {3rd}\
  ed.,\ \bibinfo {series} {Pure and Applied Mathematics}, Vol.\ \bibinfo
  {volume} {142}\ (\bibinfo  {publisher} {Elsevier},\ \bibinfo {address} {New
  York},\ \bibinfo {year} {2004})\BibitemShut {NoStop}%
\end{thebibliography}%

\end{document}